\documentclass[%
 reprint,
superscriptaddress,
nofootinbib,
longbibliography,
bibnotes,
 amsmath,amssymb,
 aps,
pre,
]{revtex4-1}

\usepackage{graphicx}
\usepackage{dcolumn}
\usepackage{bm}
\usepackage{color}
\definecolor{OliveGreen}{rgb}{0,0.6,0}
\definecolor{auburn}{rgb}{0.43, 0.21, 0.1}
\definecolor{blue_violet}{rgb}{0.54, 0.17, 0.89}
\definecolor{hokie_maroon}{RGB}{99, 0, 49} 
\definecolor{hokie_orange}{RGB}{207, 69, 32} 

\usepackage{appendix}
\usepackage{chngcntr}
\usepackage{etoolbox}
\usepackage{lipsum}

\AtBeginEnvironment{appendices}{%
\addcontentsline{toc}{section}{Appendices}
\counterwithin{figure}{section}
\counterwithin{equation}{section}
\counterwithin{table}{section}
}

\begin{document}


\title{Low-temperature Atomistic Spin Relaxation and Non-equilibrium Intensive Properties Using Steepest-Entropy-Ascent Quantum-Inspired Thermodynamics Modeling }

\author{Ryo Yamada}
\email{ryo213@vt.edu}
\thanks{Present address: Division of Materials and Manufacturing Science, Graduate School of Engineering, Osaka University, Suita, Osaka 565-0871, JAPAN}
\affiliation{Materials Science and Engineering Department, Virginia Polytechnic Institute and State University, Blacksburg, Virginia 24061, USA}
\author{Michael R. von Spakovsky}
\email{vonspako@vt.edu}
\affiliation{Center for Energy Systems Research, Mechanical Engineering Department, Virginia Polytechnic Institute and State University, Blacksburg, Virginia 24061, USA}
\author{William T. Reynolds, Jr.}
\email{reynolds@vt.edu}
\affiliation{Materials Science and Engineering Department, Virginia Polytechnic Institute and State University, Blacksburg, Virginia 24061, USA}

\date{\today}

\begin{abstract}
The magnetization of body-centered cubic iron at low temperatures is calculated with the steepest-entropy-ascent quantum thermodynamics (SEAQT) framework. This framework assumes that a thermodynamic property in an isolated system traces the path through state space with the greatest entropy production. Magnetization is calculated from the expected value of a thermodynamic ensemble of quantized spin waves based on the Heisenberg spin model applied to an ensemble of coupled harmonic oscillators. A realistic energy landscape is obtained from a magnon dispersion relation calculated using spin-density-functional-theory. The equilibrium magnetization as well as the evolution of magnetization from a non-equilibrium state to equilibrium are calculated from the path of steepest entropy ascent determined from the SEAQT equation of motion in state space. The framework makes it possible to model the temperature- and time-dependence of magnetization without a detailed description of magnetic damping.  The approach is also used to define intensive properties (temperature and magnetic field strength) that are fundamentally, i.e., canonically or grand canonically, valid for any non-equilibrium state. Given the assumed magnon dispersion relation, the SEAQT framework is used to calculate the equilibrium magnetization at different temperatures and external magnetic fields and the results are shown to closely agree with experiment for temperatures less than 500 K. The time-dependent evolution of magnetization from different initial states and interactions with a reservoir is also predicted.
\end{abstract}

\maketitle

\section{\label{chap4_sec:level1}Introduction}
 
Ferromagnetic materials are challenging to model because magnetic properties arise from hierarchical interactions arising from different spatial and temporal scales. In the broadest sense, magnetic moments have two possible sources: intrinsic moments associated with elementary particles and moments caused by moving charge.  Intrinsic moments are associated with the quantum descriptions of electrons and protons, while moments from moving charge arise from the collective motion of atoms and molecules within a crystal as well as the flow of electrons and ions. To determine the net magnetization from the underlying moments, different frameworks are needed to account for the interactions among these component contributions across spatial and temporal scales. 

At the smallest spatial scale, electronic structure calculations such as tight-binding and density functional theory (DFT) with spin polarization can estimate the magnetic moment of individual atoms. Electronic structure methods can be combined with a Hamiltonian derived from the Heisenberg spin model to construct atomistic models that can calculate Heisenberg exchange energies and the magnetic moments associated with small crystals \cite{evans2014atomistic}. 

For time-dependent behavior, spin dynamic models incorporate magnetic damping to enable simulations in time  \cite{evans2014atomistic,eriksson2017atomistic}. The simulations are carried out at finite temperatures (i.e., at temperatures greater than zero) because thermal fluctuations introduce stochastic torques on the magnetic moments that contribute to damping. Spin dynamic models employ an equation of motion based upon Langevin dynamics \cite{evans2014atomistic,weiss2008quantum,ma2011langevin,ma2012longitudinal} or the stochastic Landau-Lifshitz-Gilbert equation \cite{antropov1996spin,skubic2008method,eriksson2017atomistic,evans2014atomistic}.  Spin dynamics have been coupled with molecular dynamics in spin-lattice dynamic simulations \cite{ma2008large} and spin-lattice-electron dynamic simulations \cite{ma2012spin}. The latter include coupled spin, atomic, and electronic degrees of freedom. These simulations are, of course, all carried out at atomistic scales. Note also that in addition to the temperature effect on the dynamics, the magnetization is affected via the so-called magnetocaloric effect, a coupling between spin and temperature that affects thermodynamic equilibrium \cite{de2010theoretical}.

At larger scales, i.e., for crystals larger than a few tens of nanometers, micromagnetic models \cite{brown1963thermal,brown1978krieger} are available for describing the spatial distribution of magnetization as well as magnetic domain walls and their dynamic movement. Micromagnetic models employ a continuum framework with a phenomenological description of magnetic energy consisting of exchange, anisotropy, Zeeman, demagnitizing field, and magnetoelastic contributions. 

Despite the sophistication of all of these approaches, challenges remain including combining models across multiple scales and treating the influence of local defects on magnetization \cite{perera2016reinventing,weinan2011principles}. In the present contribution, a non-equilibrium thermodynamic, ensemble-based, modeling framework is described that has the potential to address these challenges. The framework is based on the unification of quantum mechanics and thermodynamics provided by steepest-entropy-ascent quantum thermodynamics (SEAQT) \cite{hatsopoulos1976-I,hatsopoulos1976-IIa,hatsopoulos1976-IIb,hatsopoulos1976-III,beretta2005generalPhD,beretta2014steepest,li2016steepest,li2016generalized,li2018interacting}. It has been applied to a variety of both quantum and classical kinetic phenomena such as entanglement and coherence \cite{cano2015steepest}, chemical reactions \cite{li2016steepest,beretta2017steepest}, the heat and mass diffusion of indistinguishable particles \cite{li2016generalized,li2017study,li2018multiscale}, the thermal expansion of silver \cite{yamada2018thermalexpansion}, and phase decomposition in alloys \cite{yamada2019continuousdiscontinous,yamada2019orderingspinodal}. This novel framework has a number of attractive features \cite{yamada2019methodology}, namely, that it is valid at any temperature, is readily scalable, can accommodate system interactions such as external fields or exchanges of heat without {\it ad hoc} modifications, and has relatively modest computational resource requirements. In the context of magnetic modeling, it utilizes an energy landscape and the SEAQT equation of motion to describe the dynamic behavior of a system through a rate of entropy production --- an easily calculated quantity within this framework that can be applied to multiple spatial and temporal scales without a mechanistic description of magnetic damping.

SEAQT is based upon a first-principles, non-equilibrium thermodynamic-ensemble approach that unifies quantum mechanics and thermodynamics into a single, self-consistent framework \cite{hatsopoulos1976-I,hatsopoulos1976-IIa,hatsopoulos1976-IIb,hatsopoulos1976-III,beretta2005generalPhD}. Its most distinctive characteristics are that the entropy is viewed as a measure of the energy load sharing among the available energy eigenlevels and that relaxation processes are determined by a unique master equation, i.e., the SEAQT equation of motion. This equation drives a relaxing system in the direction of steepest entropy ascent at each instant of time without the need for invoking the local or near-equilibrium assumption, and the quantum or classical model that it provides is valid at any temporal and spatial scale and for states even far-from-equilibrium \cite{li2018multiscale}. 

Within the SEAQT theoretical framework, Li and von Spakovsky have developed the concept of hypoequilibrium states \cite{li2016steepest,li2016generalized} (i.e., a non-equilibrium relaxation pattern) and used the concept to define ``fundamental'', i.e., canonical or grand canonical, non-equilibrium intensive properties such as temperature, chemical potential, and pressure. Using these non-equilibrium intensive properties, Li, von Spakovsky, and Hin have applied the SEAQT framework to the coupled transport of phonons and electrons, clearly distinguishing the non-equilibrium temperatures of phonons and electrons, and showed that the SEAQT equation of motion recovers the two-temperature model of electron-phonon coupling when a constant relaxation time for phonons and electrons is assumed \cite{li2018electronphonon}. They also have proven that the SEAQT formulation reduces to the Boltzmann transport equations in the near-equilibrium limit \cite{li2018electronphonon,yamada2019methodology}.

To apply the SEAQT framework to any relaxation process, one must first build an energy landscape or energy eigenstructure  --- a set of energy eigenlevels --- for the system in question. The eigenstructure can be constructed from appropriate quantum models and degrees of freedom for the relevant particles or molecules. However, complex many-body interactions between particles in a solid phase makes the use of simple quantum mechanical models difficult. In addition, the number of energy eigenlevels in the eigenstructure of solids becomes effectively infinite as the number of particles in a system approaches those of the bulk material. To avoid these difficulties, a quantum-inspired energy eigenstructure \cite{yamada2018thermalexpansion} is constructed here with the use of reduced-order models \cite{yamada2019methodology} from which a so-called ``pseudo-eigenstructure'' is developed using a density of states method \cite{li2016steepest}. The reduced-order models replace the quantum models with simple solid-state models, while the density of states method converts the infinite energy eigenlevel structure to a finite-level one. The approach used in the thermal expansion application of Yamada, von Spakovsky, and Reynolds \cite{yamada2018thermalexpansion} in which atoms are treated as coupled oscillators, provides an analogous strategy for constructing a pseudo-eigenstructure of a spin system by viewing spins as coupled oscillators. 

In the present contribution, relaxation processes of magnetization in bcc-iron with and without an external magnetic field are investigated using the SEAQT framework with a pseudo-eigenstructure constructed from the coupled spin oscillators. Fundamental definitions of non-equilibrium intensive properties (i.e., temperature and magnetic field strength) in spin systems are also proposed using the concept of hypoequilibrium states \cite{li2016steepest}. For simplicity, harmonic oscillators are used for the reduced-order model and the focus is on magnetization in the low-temperature region (i.e., below one half the Curie temperature) where spin dynamics simulations have not succeeded in accurately predicting the magnetization. 

This paper is organized as follows. The SEAQT equation of motion in an external magnetic field and non-equilibrium intensive properties are described in Sec.\;\ref{chap4_sec:level2_1}, and the pseudo-eigenstructure is constructed assuming coupled harmonic oscillators in Sec.\;\ref{chap4_sec:level2_2}. In Section\;\ref{chap4_sec:level3}, the calculated equilibrium magnetizations for various external magnetic field strengths and temperatures are shown and compared with experimental data (Sec.\;\ref{chap4_sec:level3_1}). Some calculated relaxation processes of magnetizations are then shown in Secs.\;\ref{chap4_sec:level3_2} and \ref{chap4_sec:level3_3} focusing on relaxations in the far-from-equilibrium region and the use of non-equilibrium intensive properties, respectively. At the end, results of the magnetization calculations using the SEAQT model are summarized in Sec.\;\ref{chap4_sec:level4}.

\section{\label{chap4_sec:level2}Theory}

\subsection{\label{chap4_sec:level2_1}SEAQT equation of motion}
\subsubsection{\label{chap4_sec:level2_1_1}An isolated system}
Relaxation under the SEAQT framework arises from a fundamental equation of motion, which uses the maximum rate of entropy production principle. 
The SEAQT equation of motion is given as \cite{beretta1985quantum,beretta2006nonlinear,beretta2009nonlinear}
\begin{equation}
\frac{d\hat{\rho}}{dt}=\frac{1}{i\hbar}[\hat{\rho},\hat{H}]+\frac{1}{\tau(\hat{\rho})}\hat{D}(\hat{\rho}) \; , \label{eq4:equation_of_motion}
\end{equation}
where $\hat{\rho}$ is the density operator, $t$ the time, $\hbar$ the modified Planck constant, $\hat{H}$ the Hamiltonian operator, $\tau$ the relaxation time (which represents the rate at which the system moves along the unique thermodynamic path in Hilbert space predicted by the equation of motion), and $\hat{D}$ the dissipation operator. The left-hand side of the equation and the first term on the right corresponds to the time-dependent von Neumann equation (or time-dependent Schr\"{o}dinger equation). The second term on the right is a dissipation term, an irreversible contribution that accounts for relaxation processes in the system. The first term on the right accounts for the reversible symplectic dynamics \cite{beretta2014steepest} (e.g., quantum correlations and/or coherences). However, for many classical systems, $\hat{\rho}$ is diagonal in the Hamiltonian eigenvector basis so that $\hat{\rho}$ and $\hat{H}$ commute, and the first term on the right disappears \cite{li2016generalized,li2016modeling,li2017study}. In this case, the model is simply ``quantum-inspired''. When the only generators of the motion are the identity and Hamiltonian operators, Eq.\;(\ref{eq4:equation_of_motion}) simplifies to \cite{beretta2006nonlinear,beretta2009nonlinear,li2016steepest}
\begin{equation}
\frac{dp_j}{dt}=\frac{1}{\tau}\frac{\begin{vmatrix} 
-p_j \mathrm{ln} \frac{p_j}{g_j} & p_j & \epsilon_jp_j \\
\langle s \rangle & 1 & \langle e \rangle \\
\langle es \rangle & \langle e \rangle & \langle e^2 \rangle
\end{vmatrix}}{\begin{vmatrix} 
1 & \langle e \rangle \\
\langle e \rangle & \langle e^2 \rangle 
\end{vmatrix}} \; ,  \label{eq4:equation_of_motion_simplified}
\end{equation}
where 
\[
\begin{array}{c c}
\langle s \rangle = - \sum\limits_{i} p_i \mathrm{ln} \frac{p_i}{g_i}  \; ,
&
\langle e \rangle = \sum\limits_{i} \epsilon_i p_i  \; , \\ \\
\langle e^2 \rangle = \sum\limits_{i} \epsilon_i^2 p_i \; , \;\;\;
&
\langle es \rangle = - \sum\limits_{i} \epsilon_i p_i \mathrm{ln} \frac{p_i}{g_i}  \; .
\end{array}
\]
The $p_j$ are the diagonal terms of $\hat{\rho}$, each of which represents the occupation probability in the $j^{th}$ energy level, ${\epsilon}_j$, and the $g_j$ are the degeneracies of the energy eigenlevel. Note that the von Neumann expression for entropy is used here. Provided the density operator is based on a homogeneous ensemble, this expression satisfies all the characteristics of the entropy required by thermodynamics without making entropy a statistical property of the ensemble \cite{gyftopoulos1997entropy,cubukcu1993thermodynamics}. It is assumed that $\hat{\rho}$ is diagonal in the eigenvector basis throughout all of the calculations. This is the case when there are no quantum correlations between particles (or spins) and for many classical systems \cite{li2016generalized,li2016modeling,li2017study}.

The SEAQT equation of motion, Eq.\;(\ref{eq4:equation_of_motion_simplified}), is derived via a constrained gradient in Hilbert space that causes the system to follow the direction of steepest entropy ascent when energy and occupation probabilities are conserved (i.e., an isolated sysetm). When another conservation constraint or generator of the motion is imposed, in this case that for the magnetization, the equation of motion becomes \cite{yamada2019methodology}
\begin{equation}
\frac{dp_j}{dt}=\frac{1}{\tau}\frac{\begin{vmatrix} 
-p_j \mathrm{ln} \frac{p_j}{g_j} & p_j & \epsilon_j p_j  & m_j p_j  \\
\left< s \right> & 1 & \left< e \right> & \left< m \right> \\
\left< es \right> & \left< e \right> & \left< e^2 \right> & \left< em \right> \\
\left< ms \right> & \left< m \right> & \left< em \right> & \left< m^2 \right> \\
\end{vmatrix}}{\begin{vmatrix} 
1 &  \left< e \right> & \left< m \right> \\
 \left< e \right> & \left< e^2 \right> & \left< em \right> \\
 \left< m \right> & \left< em \right> & \left< m^2 \right> \\
\end{vmatrix}} \; ,  \label{eq4:equation_of_motion_magnetization}
\end{equation}
where 
\[
\begin{array}{c c}
\langle m \rangle = \sum\limits_{i} m_i p_i  \; ,
&
\langle m^2 \rangle = \sum\limits_{i} m_i^2 p_i \; , \\ \\
\langle em \rangle = \sum\limits_{i} \epsilon_i m_i p_i \; , \;\;\;
&
\langle ms \rangle = - \sum\limits_{i} m_i p_i \mathrm{ln} \frac{p_i}{g_i}  \; ,
\end{array}
\]
and $m_j$ is the magnetization in the $j^{th}$ energy eigenlevel. Equation\;(\ref{eq4:equation_of_motion_magnetization}) is the generalized equation of motion of Eq.\;(\ref{eq4:equation_of_motion_simplified}) for magnetic materials in an isolated system and has a form similar to one derived for the conservation of the number of particles given in Ref.\;\cite{li2018interacting}.

\subsubsection{\label{chap4_sec:level2_1_2}A composite system (two interacting systems) }
The SEAQT equations of motion, Eqs.\;(\ref{eq4:equation_of_motion_simplified}) and (\ref{eq4:equation_of_motion_magnetization}), are valid for an isolated system. The equation of motion for interacting systems (non-isolated) can be derived by treating interacting systems as an isolated composite system made up of interacting subsystems. In addition, a SEAQT equation of motion for a system interacting with a reservoir can be derived by taking one of the subsystems in the composite system as a reservoir. This strategy has been developed by Li and von Spakovsky for heat and mass interactions \cite{li2016steepest,li2016generalized,li2018interacting}; the methodology is applied here to formulate an equation of motion for subsystems undergoing heat and magnetic field interactions.

The SEAQT equation of motion for a composite isolated system of two interacting subsystems (subsystems\;$A$ and $B$) with the conservation of energy and magnetization in the composite system is given for subsystem\;$A$ by    
\begin{equation}
\frac{dp^A_j}{dt}=\frac{1}{\tau}\frac{\begin{vmatrix} 
-p^{A}_j \mathrm{ln} \frac{p^A_j}{g^A_j} & p^A_j & 0 & \epsilon^A_j p^A_j  & m^A_j p^A_j  \\
\left< s \right>^A & 1 & 0 & \left< e \right>^A & \left< m \right>^A \\
\left< s \right>^B & 0 & 1 & \left< e \right>^B & \left< m \right>^B \\
\left< es \right> & \left< e \right> ^A & \left< e \right> ^B & \left< e^2 \right> & \left< em \right> \\
\left< ms \right> & \left< m \right>^A & \left< m \right>^B & \left< em \right> & \left< m^2 \right> \\
\end{vmatrix}}{\begin{vmatrix} 
1 & 0 & \left< e \right>^A & \left< m \right>^A \\
0 & 1 & \left< e \right>^B & \left< m \right>^B \\
 \left< e \right>^A & \left< e \right>^B & \left< e^2 \right> & \left< em \right> \\
 \left< m \right>^A & \left< m \right>^B & \left< em \right> & \left< m^2 \right> \\
\end{vmatrix}} \; ,  \label{eq4:equation_of_motion_magnetization_interacting}
\end{equation}
where $\left< \cdot \right>^{A(B)}$ is the expectation value of a property in subsystem\;$A$ (or $B$), and $\left< \cdot \right>= \left< \cdot \right>^A + \left< \cdot \right>^B$ corresponds to the property of the composite system. An expression similar to Eq.\;(\ref{eq4:equation_of_motion_magnetization_interacting}) can be written for subsystem\;$B$. Whereas energy and magnetizaiton are conserved within the composite system, the occupation probabilities are conserved within each subsystem. Using the cofactors $C_1$, $C^A_2$, $C_3$, and $C_4$ of the first line of the determinant in the numerator,  Eq.\;(\ref{eq4:equation_of_motion_magnetization_interacting}) can be expressed as
\begin{equation}
\begin{split}
& \;\;\;\;\;  \frac{dp^A_j}{dt} = \frac{1}{\tau} p^A_j ( - \mathrm{ln} \frac{p^A_j}{g^A_j} - \frac{C_2^A}{C_1} - \epsilon^A_j  \frac{C_3}{C_1} + m^A_j \frac{C_4}{C_1} )  \\
& =\frac{1}{\tau} p^A_j \left[ (s^A_j -  \left< s \right>^A)  - (\epsilon^A_j - \left< e \right>^A) \frac{C_3}{C_1} + (m^A_j -  \left< m \right>^A)\frac{C_4}{C_1}  \right] \\
& \; =\frac{1}{\tau} p^A_j \left[ (s^A_j -  \left< s \right>^A)  - (\epsilon^A_j - \left< e \right>^A) \beta - (m^A_j -  \left< m \right>^A)\gamma  \right]  \; , \label{eq4:equation_of_motion_magnetization_interacting_simplified}
\end{split}
\end{equation}
where $\beta$ and $\gamma$ are defined as $\beta \equiv C_3/C_1$ and $\gamma \equiv - C_4/C_1$. The $\beta$ and $\gamma$ quantities are intensive properties related to the temperature, $T$, and the magnetic field strength, $H$, i.e., $\beta=1/k_BT$ and $\gamma=H/k_BT$ (where $k_B$ is the Boltzmann constant). 

The intensive properties, $\beta$ and $\gamma$, depend on the mole fractions of each subsystem \cite{li2016generalized}. When subsystem\;$B$ is much larger than subsystem\;$A$ (i.e., when subsystem\;$B$ is taken to be a large reservoir, $R$), these intensive properties can be denoted by $\beta^R$ and $\gamma^R$, and Eq.\;(\ref{eq4:equation_of_motion_magnetization_interacting_simplified}) is transformed into
\begin{equation}
\frac{dp_j}{dt}=\frac{1}{\tau} p_j \left[ \left( s_j - \langle s \rangle \right)  - \left( \epsilon_j - \left< e \right> \right) \beta^R - \left( m_j- \langle m \rangle \right) \gamma^R  \right]  ,     
\label{eq4:equation_motion_magnetization_heat}
\end{equation}
where $\beta^R=1/k_BT_R$, $\gamma^R=H_R/k_BT_R$, and $T_R$ and $H_R$ are, respectively, the temperature of the reservoir and the external magnetic field strength. In this context, the $A$ superscripts have been dropped in Eq.\;(\ref{eq4:equation_motion_magnetization_heat}).

\subsubsection{\label{chap4_sec:level2_1_3}Non-equilibrium intensive properties }
The concept of hypoequilibrium states developed by Li and von Spakovsky \cite{li2016steepest,li2016generalized} in the SEAQT framework permits one to define intensive properties (e.g., temperature, pressure, and chemical potential) in the non-equilibrium realm. Here, the concept is briefly described and non-equilibrium temperature and magnetic field strength are defined (details are found in Refs.\;\cite{li2016steepest,li2016generalized,li2018interacting}). 

The concept of hypoequilibrium states is based on the idea that any non-equilibrium state of a system can be described by an $M^{th}$-order hypoequilibrium state by dividing the system's state space into $M$ subspaces, each of which is described by a canonical distribution. Note that in some cases such as the one presented here, the subspaces coincide with a physical division of the system into subsystems. Furthermore, as proven in Refs.\;\cite{li2016steepest,li2018interacting}, once in a $M^{th}$-order hypoequilibrium state, the system remains in such a state throughout the entire kinetic evolution process predicted by the SEAQT equation of motion, and the relaxation paths can be simply described by the time dependence of the intensive properties in each subspace. This means that the time evolution of the probability distribution in subspace (or subsystem) $M$ can be described by 
\begin{equation}
p^{(M)}_{j}(t)= \frac{ p^{(M)}(t) }{Z^{(M)}(t)} g^{(M)}_{j} e^{- \epsilon^{(M)}_{j} \beta^{(M)}(t)  + m^{(M)}_{j} \gamma^{(M)}(t) } \; ,  \label{eq4:Mth_canonical_distribution}
\end{equation}
where $\epsilon^{(M)}_{j}$, $g^{(M)}_{j}$, and $m^{(M)}_{j}$ are, respectively, the energy eigenlevel, energy degeneracy, and magnetization in subsystem $M$, $p^{(M)}$ is the total occupation probability in the subsystem, and $Z^{(M)}(t)$ is the partition function defined as
\begin{equation}
Z^{(M)}(t) \equiv \sum_i g^{(M)}_{i} e^{- \epsilon^{(M)}_{i} \beta^{(M)}(t)  + m^{(M)}_{i} \gamma^{(M)}(t)  } \; .  \label{eq4:partition_function}
\end{equation}
In Eq.\;(\ref{eq4:Mth_canonical_distribution}), the eigenstructure of each subsystem is invariant with time, but the intensive properties, $\beta^{(M)}$ and $\gamma^{(M)}$, are time-dependent. During the evolution process, the probability distribution in each subsystem evolves together with those of the other subsystems until they reach a mutual equilibrium state with each other, and the system ends up in a stable equilibrium state (which corresponds to $M=1$). 
 
The concept of hypoequilibrium states is applied to the composite system considered here. Thus, the time-evolution of the probability distributions in the two interacting subsystems, $A$ and $B$, take the following form: 
\begin{equation}
p^{A(B)}_{j}(t)=\frac{1}{Z^{A(B)}(t)} g^{A(B)}_{j} e^{- \epsilon^{A(B)}_{j} \beta^{A(B)}(t)  + m^{A(B)}_{j} \gamma^{A(B)}(t)} \; .  \label{eq4:second_order_hypoequilibrium_time_evolution}
\end{equation}
The time evolution of the intensive properties are determined from the equation of motion for the intensive properties \cite{li2018interacting} given by 
\begin{equation}
\begin{split}
\frac{d\beta^{A(B)}(t)}{dt} & =-\frac{1}{\tau} \left(\beta^{A(B)}(t) - \beta(t) \right) \\
\frac{d\gamma^{A(B)}(t)}{dt} & =-\frac{1}{\tau} \left( \gamma^{A(B)}(t) - \gamma(t) \right)  \; ,  \label{eq4:equation_of_motion_intensive_properties}
\end{split}
\end{equation}
where $\beta(t)=C_3(t)/C_1(t)$ and $\gamma(t)=-C_4(t)/C_1(t)$ (as defined in Eq.\;(\ref{eq4:equation_of_motion_magnetization_interacting_simplified})). 
Therefore, using Eq.\;(\ref{eq4:equation_of_motion_intensive_properties}) with Eq.\;(\ref{eq4:second_order_hypoequilibrium_time_evolution}), kinetic trajectories of the relaxation process in each subsystem can be described with non-equilibrium intensive properties (i.e., the temperature and the magnetic field strength).  In contrast to other possible definitions of non-equilibrium temperature for a spin-system \cite{ma2010temperature}, this intensive temperature is a fundamental property analogous to the temperature defined at equilibrium via a canonical distribution.

Finally, it should be noted that the hypoequilibrium concept is well-defined for any state even one far from equilibrium since the order of $M$ can be as high as needed to adequately describe the state. Furthermore, this concept --- unlike the local-equilibrium assumption of continuum mechanics --- is fundamental and does not rely on the assumption that the total system must be subdivided into a set of infinitesimally small local systems each of which is assumed phenomenologically to be in stable equilibrium due to the smallness of the gradients across it. In fact, neither size nor gradients nor for that matter a physical division of the system are necessary requirements for the hypoequilibrium description.

\subsection{\label{chap4_sec:level2_2}Pseudo-eigenstructure}
From the Heisenberg Hamiltonian, the total energy of spin systems is given by \cite{kittel1966introduction,halilov1998adiabatic,pajda2001ab,kormann2009pressure}
\begin{equation}
E=-\sum_{ij} J_{ij} \bm{\mathrm{e}}_i \cdot \bm{\mathrm{e}}_j  \; ,     \label{eq4:Hamiltonian_of_heisenberg}
\end{equation}
where the $J_{ij}$ represent exchange interaction energies, and $\bm{\mathrm{e}}_i$ and $\bm{\mathrm{e}}_j$ are the unit vectors in the direction of the local magnetic moment at lattice sites $i$ and $j$. The magnon dispersion relation, which is directly related to the energy eigenlevels of spin waves (magnons), has been calculated from Eq.\;(\ref{eq4:Hamiltonian_of_heisenberg}) using spin-density-functional-theory\cite{halilov1998adiabatic,pajda2001ab,kormann2009pressure}, and the result calculated by Padja $et$ $al.$\cite{pajda2001ab} is used here by fitting to a function \cite{moran2003ab}:
\begin{equation}
\hbar \omega=D |\bm{\mathrm{k}}|^2+E' |\bm{\mathrm{k}}|^4+E''(k_x^2 k_y^2 + k_y^2 k_z^2 + k_x^2 k_z^2) \; ,     \label{eq4:magnetization_dispersion_function}
\end{equation}
where $\omega$ is the magnon frequency, $\bm{\mathrm{k}}$ is the wave vector, $k_x$, $k_y$, and $k_z$ are the components of $\bm{\mathrm{k}}$, and $D$, $E'$, and $E''$ are fitting parameters (shown in Table\;\ref{table4:magnon_fitting}). The spin waves have different degeneracies depending upon the frequency, $\omega$. The degeneracy (or the density of states) is given by \cite{kittel1966introduction}
\begin{equation}
g(\omega)=4\pi k^2 \left(\frac{1}{2\pi} \right) ^3 \left( \frac{dk}{d\omega} \right) ^ {\frac{3}{2}} \; ,    \label{eq4:dos_magnon}
\end{equation}
where $(d\omega /dk)^{-1}$ is calculated from Eq.\;(\ref{eq4:magnetization_dispersion_function}).
\begin{table}
\begin{center}
\caption{\label{table4:magnon_fitting} The fitting parameters in Eq.\;(\ref{eq4:magnetization_dispersion_function}) for the magnon dispersion relation \cite{pajda2001ab}. Experimental data for the atomic volume of bcc-Fe at the ground state, $V=11.81$ $(\mbox{\AA}^3 /\mbox{atom})$ \cite{kittel1966introduction}, is assumed here. }
\footnotesize
\begin{tabular}{ c  c  }
$\quad$ & $\quad$ \\   \hline \hline
$\quad\quad$  &  \quad $V=11.81 \; (\mbox{\AA}^3 /\mbox{atom}) $ \quad   \\ \hline
\;$D$ (meV\;\AA) & 247.7  \\
\;$E'$ (meV\;\AA$^4$) & -31.73  \\
\;$E''$ (meV\;\AA$^4$) & 1.970  \\ \hline \hline
\end{tabular}
\end{center}
\end{table}

The eigenenergy of each magnon, $\epsilon_n$, is \cite{kittel1966introduction}
\begin{equation}
\epsilon_n=\left( n+\frac{1}{2} \right) \hbar\omega  \; ,  \label{eq4:eigenvalue_magnon}
\end{equation}
where $n$ is the quantum number ($n=0,1,2,...$). Incorporation of the magnon frequency, Eq.\;(\ref{eq4:magnetization_dispersion_function}), into Eq.\;(\ref{eq4:eigenvalue_magnon}) cannot be done in a straightforward manner because the density of states, Eq.\;(\ref{eq4:dos_magnon}), is not discrete. To circumvent this difficulty, the density of states method developed by Li and von Spakovsky \cite{li2016steepest} within the SEAQT framework is applied (see also Ref.\;\cite{yamada2018thermalexpansion} for a similar application of the method). Using this approach, the magnon frequency and degeneracy in the system become 
\begin{equation}
\omega_j=\frac{1}{G_j}\int_{\bar{\omega}_j}^{\bar{\omega}_{j+1}} \bar{\omega} g (\bar{\omega}) \; d\bar{\omega}   \label{eq4:magnonl_frequency_pseud}
\end{equation}
and
\begin{equation}
G_j=\int_{\bar{\omega}_j}^{\bar{\omega}_{j+1}} g(\bar{\omega}) \; d\bar{\omega} \; ,  \label{eq4:magnon_degeneracy_pseud}
\end{equation}
The quantity, $\bar{\omega}_j$, is the $j^{th}$ frequency interval in the original infinite energy-eigenlevel system determined by
\begin{equation}
\bar{\omega}_j= (j-1) \Delta \bar{\omega}=(j-1)\frac{\bar{\omega}_{\mbox{\scriptsize cut}}}{R} \; ,  \label{eq4:magnon_intervals}
\end{equation}
where $R$ is the number of intervals, $j$ is an integer ($j=0,1,2,...R$), and $\bar{\omega}_{\mbox{\scriptsize cut}}$ is the cutoff frequency estimated by using $|\bm{\mathrm{k}}|=2\pi/a$ (where $a$ is the lattice constant) in Eq.\;(\ref{eq4:magnetization_dispersion_function}). Using Eqs.\;(\ref{eq4:eigenvalue_magnon}) and (\ref{eq4:magnonl_frequency_pseud}), the energy eigenlevels become
\begin{equation}
E_{j,n}=\left( n+\frac{1}{2} \right) \hbar\omega_j \;  . \label{eq4:magnon_energy_pseud}
\end{equation}
Furthermore, the magnetization (change) of the energy eigenlevel is given as 
\begin{equation}
M_{j,n}= - n \mu  \; , \label{magnon_magnetization_pseud}
\end{equation}
where $\mu$ is the magnetic moment of iron ($\mu=2.22\mu_B$ where $\mu_B$ is the Bohr magneton \cite{kittel1966introduction}) and the magnetization at the ground state ($n=0$) is taken as ``zero''. Henceforth, the occupation probability, energy eigenlevel, energy degeneracy, and magnetization are denoted $P_{j,n}$, $E_{j,n}$, $G_{j,n}$, and $M_{j,n}$, respectively (instead of as $p_{j,n}$, $\epsilon_{j,n}$, $g_{j,n}$, and $m_{j,n}$) in order to emphasize these quantities apply to the finite energy-eigenlevel system. Note that each frequency interval, $\omega_j$, has the same energy degeneracy; that is, $G_{j,n}=G_{j}$.

The decrease in magnetization caused by non-aligned spins is given by a weighted average of the expected magnetizations in each frequency \cite{kittel1966introduction}. Therefore, the magnetization, $M$, in a system is
\begin{equation}
M= \mu (1-  \sum_i G_{i} \langle M \rangle_i )   \; ,     \label{eq4:magnetization_magnon}
\end{equation}
where the expectation magnetization of each magnon frequency, $ \langle M \rangle_i $, is
\begin{equation}
\langle M \rangle_i  =  \sum_{n} M_{i,n} P'_{i,n} =  \sum_{n} M_{i,n} \frac{ P_{i,n} }{\sum_{n'}  P_{i,n'} }  \; ,     \label{eq4:magnon_number_average}
\end{equation}
where $P'_{i,n}$ is the normalized occupation probability at each magnon frequency.

Note that longitudinal degrees of freedom of atomic magnetic moments (or the magnitude of spins) and anharmonic effects (or magnon-magnon interactions) are ignored for the sake of simplicity. While the former is not sensitive to temperature in bcc-iron below the Curie temperature $T_c$~ \cite{ma2012longitudinal}, the latter effect only becomes significant above about half the Curie temperature. Anharmonic effects could be included using anharmonic oscillators as was done for analogous thermal expansion calculations in Ref.\;\cite{yamada2018thermalexpansion}. However, that added complexity is beyond the scope of this paper, and the validity of the present approach is limited to low temperatures, $T < T_c/2$, where anharmonic effects can be neglected.

\section{\label{chap4_sec:level3}Results and Discussion}

\subsection{\label{chap4_sec:level3_1}Equilibrium magnetization}
The equilibrium magnetization at different temperatures and external magnetic fields is determined from the following canonical distribution:
\begin{equation}
P^{\mbox{\footnotesize  se}}_{k}=\frac{G_{k} e^{- \left( E_{k} - M_{k} H^{\mbox{\footnotesize  se}} \right) /k_BT^{\mbox{\footnotesize  se}}}}{Z^{\mbox{\footnotesize  se}}} \; ,  \label{eq4:canonical_distribution}
\end{equation}
where $Z^{\mbox{\footnotesize  se}}$ is the partition function already defined in Eq.\;(\ref{eq4:partition_function}), ``$\mbox{se\/}$'' denotes stable equilibrium, and the subscripts corresponding to each energy eigenlevel are simply designated by a $k$ (e.g., $E_{j,n} \rightarrow E_k $). 

\begin{figure}
\begin{center}
\includegraphics[scale=0.48]{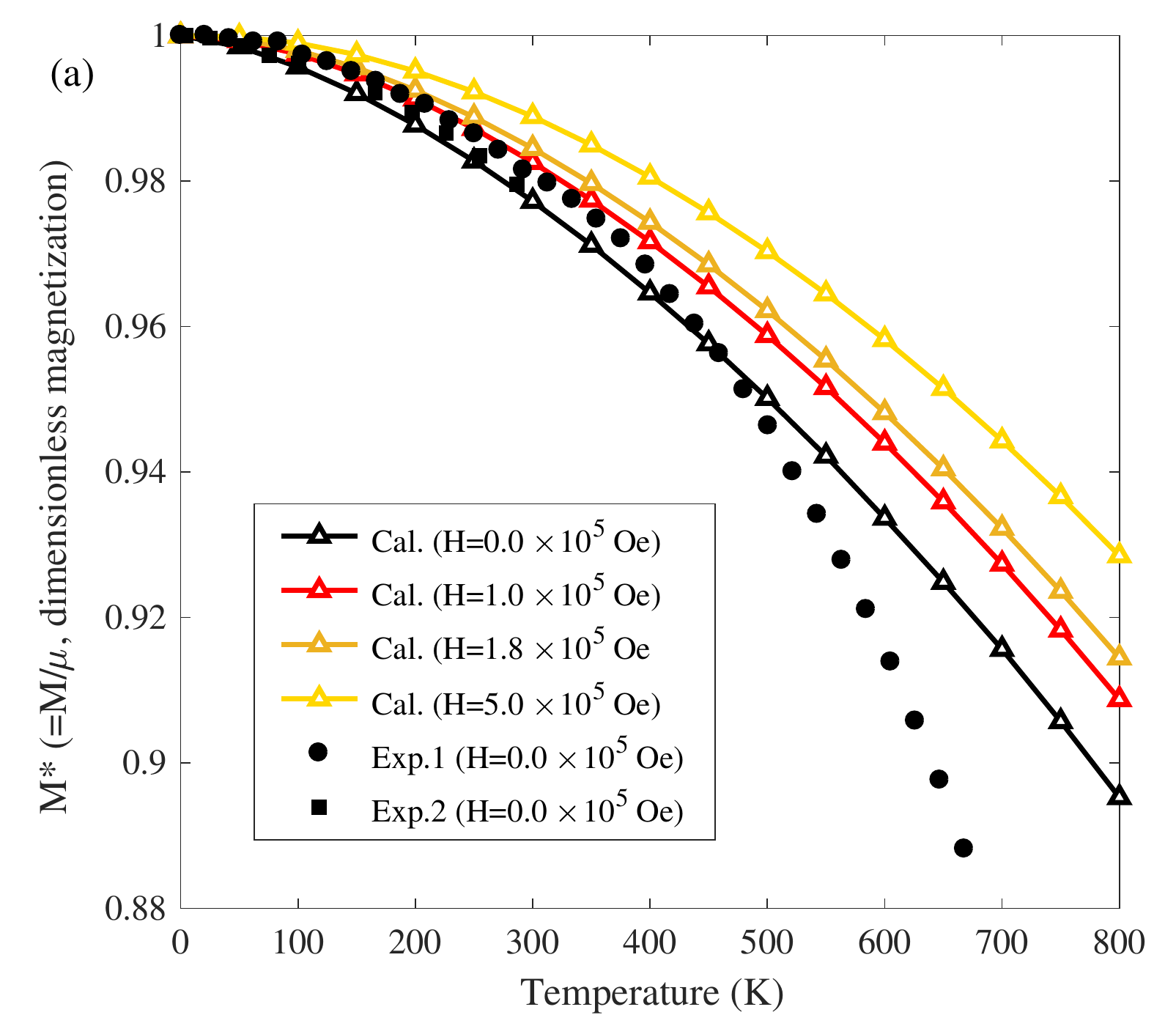}
\includegraphics[scale=0.48]{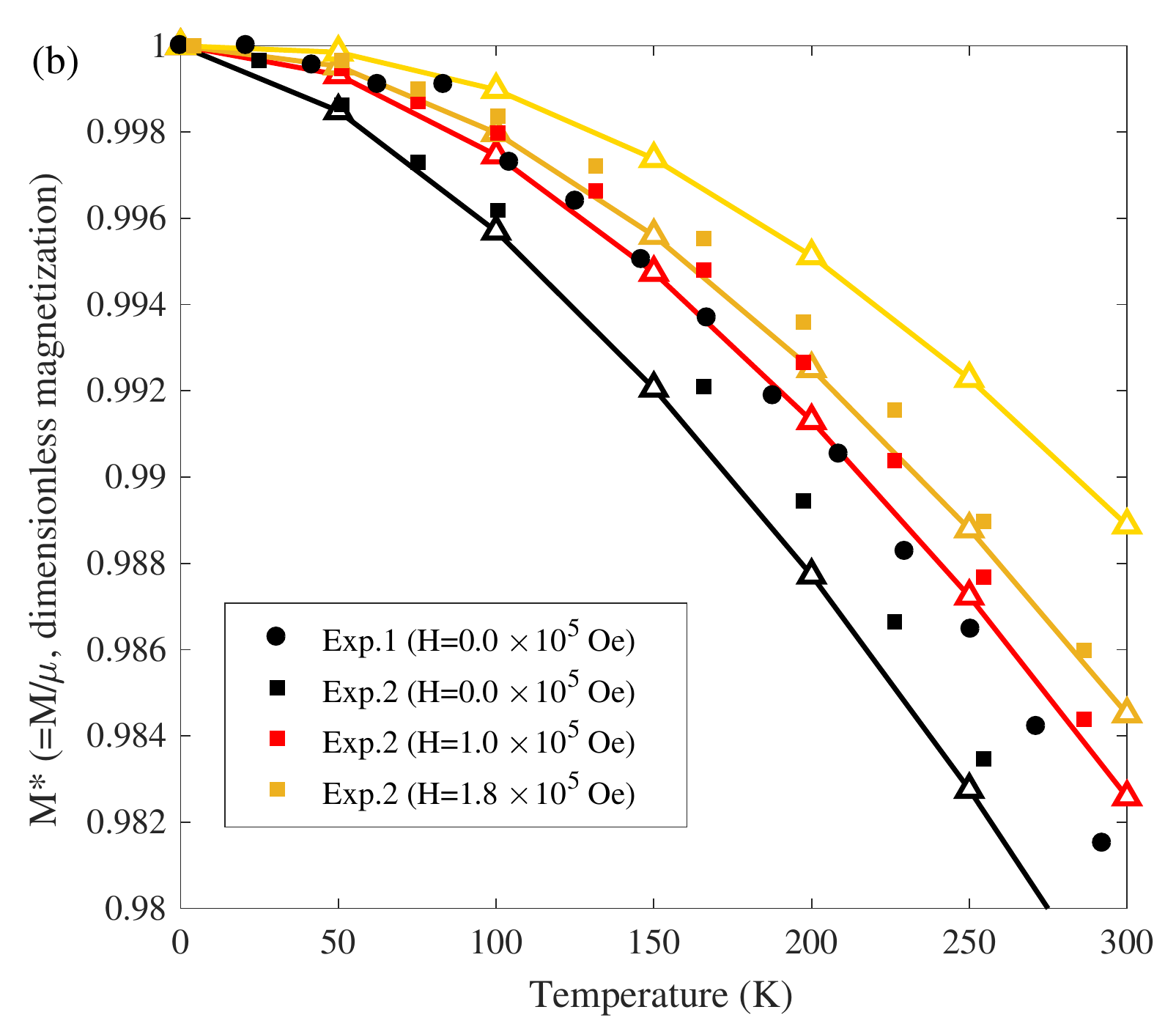}
\caption{\label{fig4:normalized_magnetization_spin_wave} The calculated temperature dependence of equilibrium magnetization at various external magnetic fields; (b) is an enlarged portion of (a) below room temperature. The experimental data are shown as solid circles \cite{crangle1971magnetization} and squares \cite{pauthenet1982spin}. The magnetization is plotted as a fraction of the magnetic moment of iron, $\mu$.}
\end{center}
\end{figure}

The calculated equilibrium magnetization, $M$, using Eqs.\;(\ref{eq4:magnetization_magnon}) and (\ref{eq4:canonical_distribution}) is shown in Fig.\;\ref{fig4:normalized_magnetization_spin_wave} where the magnetization is expressed as a fraction of the magnetic moment of iron, $\mu$. It can be seen that the calculated results are close to the experimental data at low temperatures ($T<500$\;K) and reproduce the dependence on external magnetic fields. However, the calculated magnetization deviates from the experiments at high temperatures. As noted above, this is a consequence of anharmonic effects (magnon-magnon interactions) that are not taken into account in the model used here, i.e., coupled harmonic oscillators. If anharmonic oscillators were used, it is expected that the energy eigenlevels with large magnon quantum numbers would be lowered (see Eq.\;(\ref{eq4:eigenvalue_magnon}) or (\ref{eq4:magnon_energy_pseud})) and the equilibrium magnetizations at high temperatures would be more accurate.  

\subsection{\label{chap4_sec:level3_2}Relaxation far from equilibrium}
\begin{figure}
\begin{center}
\includegraphics[scale=0.45]{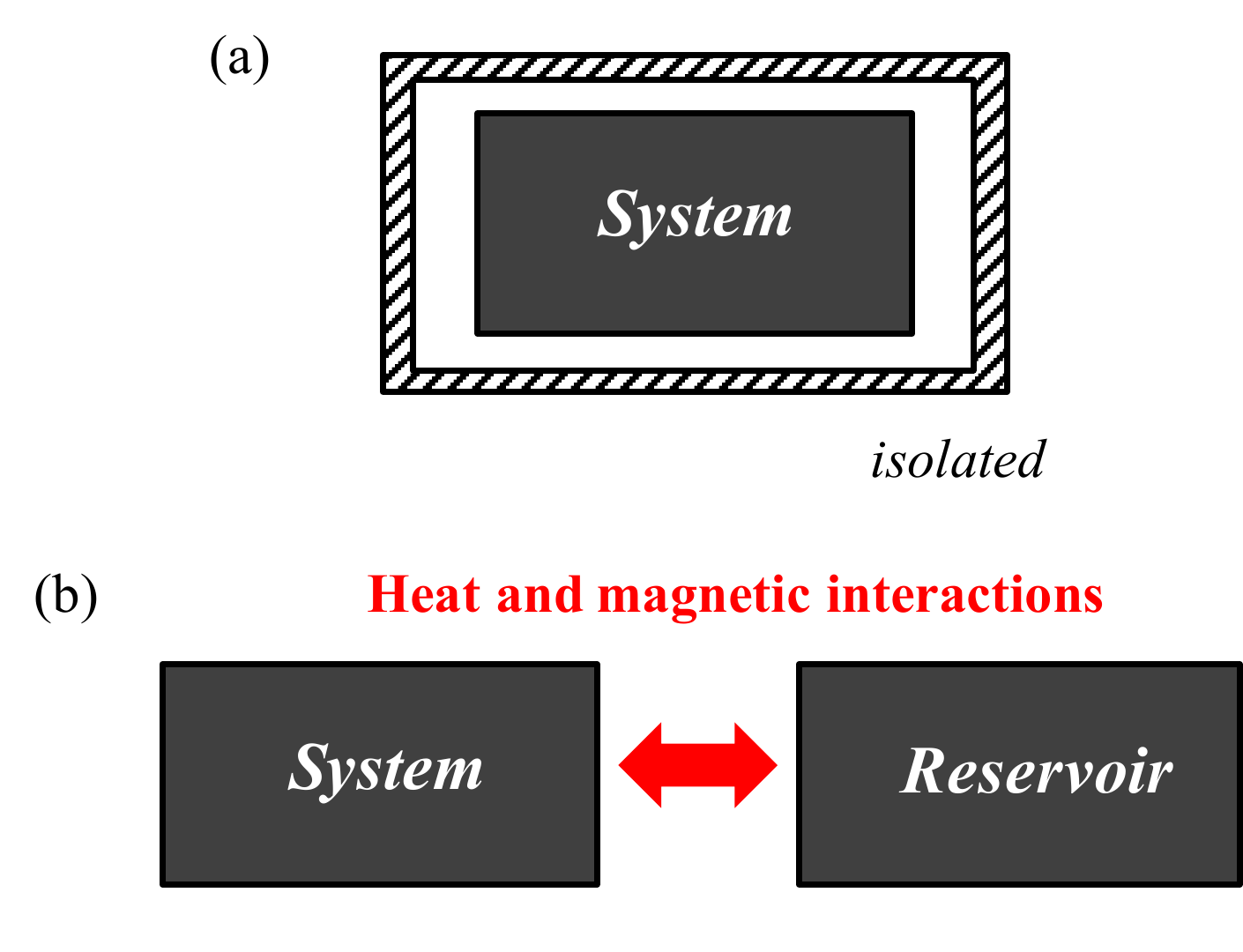}
\caption{\label{fig4:assumed_system} The system descriptions considered in Sec.\;\ref{chap4_sec:level3_2}. (a) is a depiction of the isolated system associated with the relaxation process whose initial state is based on the perturbation of a partially canonical equilibrium (P.E.) state, while (b) is a depiction of the system interacting with a reservoir associated with the relaxation process whose initial state is generated from a gamma function distribution. }
\end{center}
\end{figure}

One of the advantages of the SEAQT model is that a relaxation process to stable equilibrium from any initial non-equilibrium state (not only near equilibrium but also far from equilibrium) can be calculated without any unphysical assumptions. Relaxation processes from states in the far-from-equilibrium realm are explored in this section with/without external magnetic fields. There are an infinite number of non-equilibrium states and a variety of ways to generate an initial non-equilibrium state. Following Ref.\;\cite{yamada2018thermalexpansion}, two different approaches are considered: one using a partially canonical distribution and a second based on the gamma function distribution. While the relaxation process using the former is for an isolated system, that using the latter is for a system, which interacts with a reservoir (see Fig.\;\ref{fig4:assumed_system}).

For the isolated system case, the initial probability distribution, $P^0_k$, is generated in terms of a linear perturbation function \cite{beretta2006nonlinear}:
\begin{equation}
P^0_{k}=(1-\lambda_{\mbox{\scriptsize const}})P^{\mbox{\footnotesize pe}}_{k}+\lambda_{\mbox{\scriptsize const}}P^{\mbox{\footnotesize se}}_{k} \; , \label{eq4:initial_probability_partial}
\end{equation}
where $P^{\; \mbox{\footnotesize pe/se}}_{k}$ are the partially canonical/stable equilibrium probability distributions ($P^{\; \mbox{\footnotesize se}}_{k}$ is shown in Eq.\,(\ref{eq4:canonical_distribution})) and $\lambda_{\mbox{\scriptsize const}}$ is the perturbation constant that describes the initial departure from the partially canonical state. The partially canonical distribution when there is no external magnetic field is given by
\begin{equation}
P^{\mbox{\footnotesize pe}}_{k}=\frac{ \delta_{k} G_{k} e^{-  E_{k} /k_BT^{\mbox{\footnotesize pe}}}}{\sum_{k'} \delta_{k'} G_{k'} e^{- E_{k'} /k_BT^{\mbox{\footnotesize pe}}}} \; ,
\label{eq4:partial_canonical_distribution}
\end{equation}
where $\delta_k$ takes one or zero depending upon whether the states are occupied or unoccupied, and $T^{\mbox{\footnotesize pe}}$ is determined via the relation, $\sum_{k'} P^{\mbox{\footnotesize pe}}_{k'} E_{k'}=\sum_{k'} P^{\mbox{\footnotesize se}}_{k'} E_{k'}$, which ensures a relaxation of the isolated system to a final equilibrium state with the temperature $T^{\mbox{\footnotesize se}}$. Another way to prepare a partially canonical distribution that uses different $T^{\mbox{\footnotesize pe}}$ for each magnon frequency $\omega_j$, i.e., $T^{\mbox{\footnotesize pe}}_j$, is to use the relation, $\sum_{n} P^{\mbox{\footnotesize pe}}_{j,n} E_{j,n}=\sum_{n} P^{\mbox{\footnotesize se}}_{j,n} E_{j,n}$. Here both the partially canonical and canonical distributions are employed assuming that the lowest three quantum levels (i.e., $n=$0, 1, and $2$) are not occupied. 

\begin{figure}
\begin{center}
\includegraphics[scale=0.48]{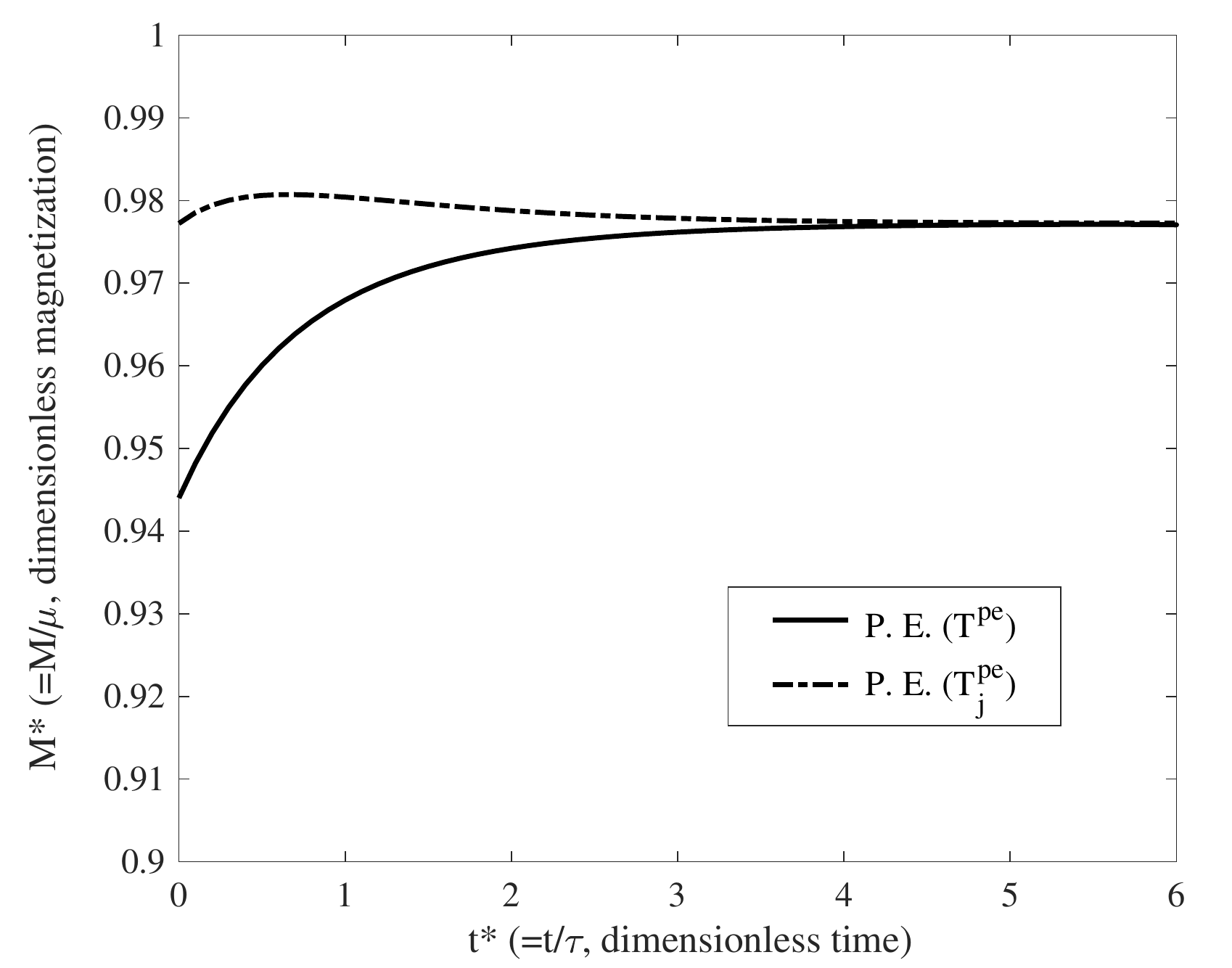}
\caption{\label{fig4:relaxation_process_partial_equilibrium} The calculated spin relaxation from two different initial non-equilibrium states prepared using the partially canonical distribution, Eq.\;(\ref{eq4:initial_probability_partial}), with $T^{\mbox{\footnotesize pe}}$ or $T^{\mbox{\footnotesize pe}}_j$. The SEAQT equation of motion for an isolated system, Eq.\;(\ref{eq4:equation_of_motion_magnetization}), is used for the kinetic calculations. The magnetization is plotted as a fraction of the magnetic moment of iron, $\mu$, and the time is normalized by the relaxation time, $\tau$.}
\end{center}
\end{figure}

The calculated spin relaxation processes from two different initial states (i.e., one based on $T^{\mbox{\footnotesize pe}}$ and the other on $T^{\mbox{\footnotesize pe}}_j$) generated using Eq.\;(\ref{eq4:initial_probability_partial}) with $\lambda=0.1$ are shown in Fig.\;\ref{fig4:relaxation_process_partial_equilibrium}. Equation\;(\ref{eq4:equation_of_motion_magnetization}) is used for the relaxation processes of the isolated system of Fig.\;\ref{fig4:assumed_system}\;(a) with $T^{\mbox{\footnotesize se}}$ set to 300\;K. It can be seen that both magnetizations relax to the equilibrium value at 300\;K with a zero external magnetic field (see Fig.\;\ref{fig4:normalized_magnetization_spin_wave}), which is independently calculated from the canonical distribution, Eq.\;(\ref{eq4:canonical_distribution}). Relaxation from an initial state prepared with $T^{\mbox{\footnotesize pe}}_j$ is particularly interesting in that the magnetization evolves non-monatonically with time.

\begin{figure}
\begin{center}
\includegraphics[scale=0.165]{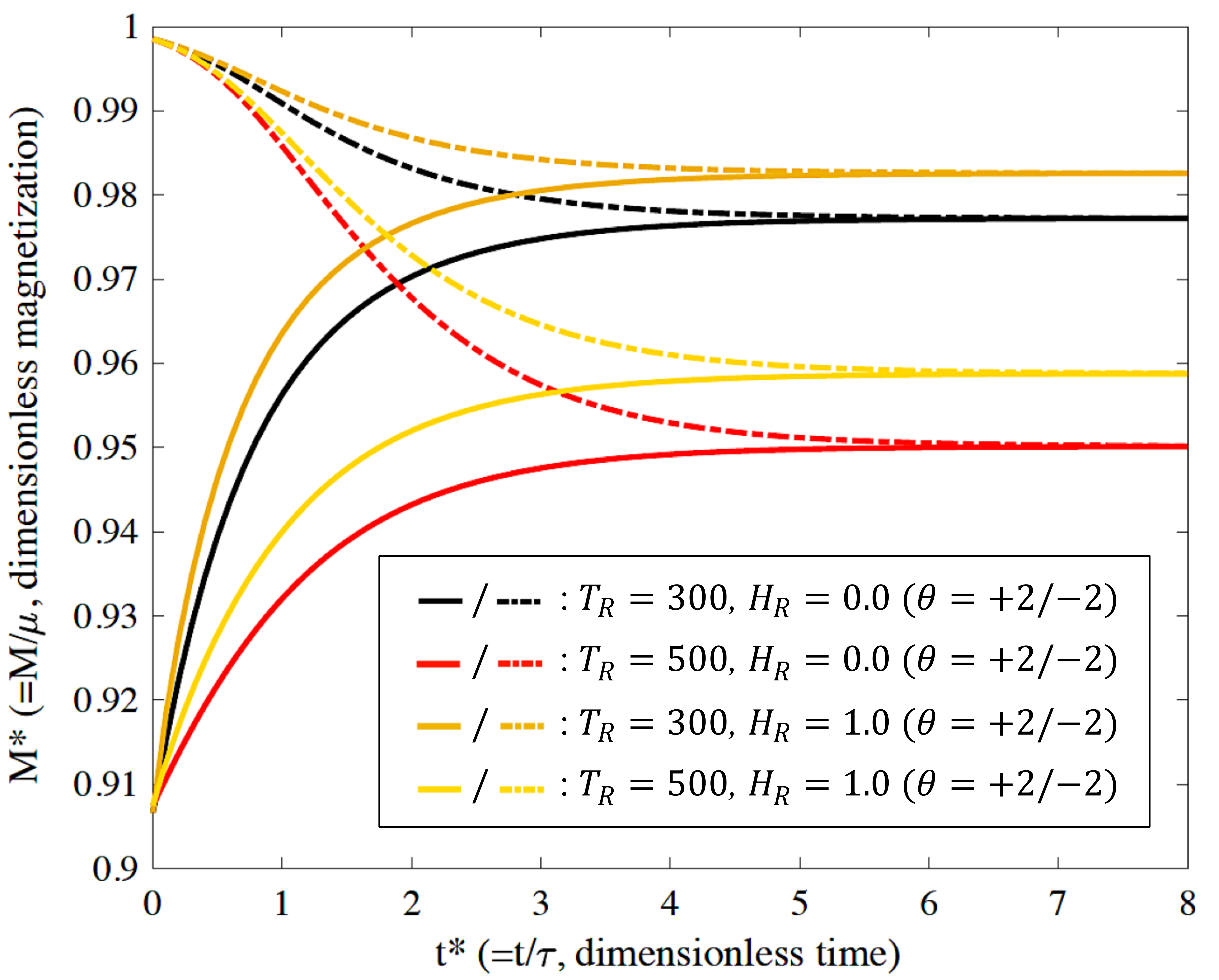}
\caption{\label{fig4:relaxation_process_gamma_distribution} The calculated relaxation processes of magnetization from two different initial states prepared using the gamma function, Eq.\;(\ref{eq4:initial_probability_gamma}), with $T_0=300$\;K, $H_0=0.0$\;Oe, and $\theta=+2$ or -2 (each of which corresponds to $M^*\approx 0.91$ or 1 at $t^*=0$; evolutions are shown in solid or broken lines, respectively). Four different combinations of reservoir temperature, $T_R$ (K), and external magnetic field strength, $H_R$ ($\times 10^5$\;Oe), are used here as indicated in the inset box. The colors represent different combinations of $T_R$ and $H_R$. The spin relaxation processes are calculated using the SEAQT equation of motion for a system interacting with a reservoir, Eq.\;(\ref{eq4:equation_motion_magnetization_heat}). The magnetization is plotted as a fraction of the magnetic moment of iron, $\mu$, and the time is normalized by the relaxation time, $\tau$.}
\end{center}
\end{figure}

In the second approach, the initial states are generated with a gamma function distribution of the form \cite{li2016steepest, li2016generalized}
\begin{equation}
P^0_{k} = \frac{G_{k} E_{k}^{\theta} e^{- \left( E_{k} - M_{k} H_0 \right) /k_BT_0}}{\sum_{k'} G_{k'} E_{k'}^{\theta}e^{- (E_{k'} - M_{k'} H_0 ) / k_BT_0}} \; , 
\label{eq4:initial_probability_gamma}
\end{equation}
where $T_0$ and $H_0$ are the initial temperature and magnetic field and $\theta$ represents an adjustable parameter that can be positive or negative and shifts the initial state away from the canonical distribution. Figure\;\ref{fig4:relaxation_process_gamma_distribution} shows the time evolutions of the magnetization for the system of Fig.\;\ref{fig4:assumed_system}\;(b) relaxing to four different stable equilibrium states (i.e., those corresponding to four sets of reservoir temperatures, $T_R$'s, and external magnetic field strengths, $H_R$'s) beginning from two different initial states generated using Eq.\;(\ref{eq4:initial_probability_gamma}) with $\theta= \pm 2.0$, $T_0=300$\;K, and $H_0=0.0$\;Oe. The relaxations are calculated using Eq.\;(\ref{eq4:equation_motion_magnetization_heat}), where the system interacts with a reservoir (see Fig.\;\ref{fig4:assumed_system}\;(b)). It can be observed that although the magnetizations of the two initial states prepared using $\theta=\pm 2.0$ are different, the final equilibrium states are same (for a given set of $T_R$ and $H_R$). The final equilibrium states correspond with the results shown in Fig.\;\ref{fig4:normalized_magnetization_spin_wave}.

One might view the approach using a partially canonical distribution as a description of spin-pumping in which applied microwave energy excites spins from low energy levels, but there is no obvious physical meaning to the initial states prepared using the gamma function distributions. They are employed here as an arbitrary means of displacing the initial non-equilibrium state far from equilibrium.

Note that the time scale in Figs.\;\ref{fig4:relaxation_process_partial_equilibrium} and \ref{fig4:relaxation_process_gamma_distribution} and the remaining figures below is normalized by the relaxation time, $\tau$. This time can be correlated with the real time, $t$, via comparisons to experimental data \cite{cano2015steepest,beretta2017steepest,li2018multiscale} or from $ab$ $initio$ calculations \cite{beretta2014steepest,li2018electronphonon,yamada2018thermalexpansion}. Real-time scaling for magnetic relaxation processes has been done in spin dynamics simulations using experimental data of the demagnetization on iron thin films induced by a laser pulse \cite{ma2012spin,ma2012longitudinal}. Although it was not attempted here, the same strategy could be taken.

\subsection{\label{chap4_sec:level3_3}Relaxation and non-equilibrium intensive properties}
\begin{figure}
\begin{center}
\includegraphics[scale=0.45]{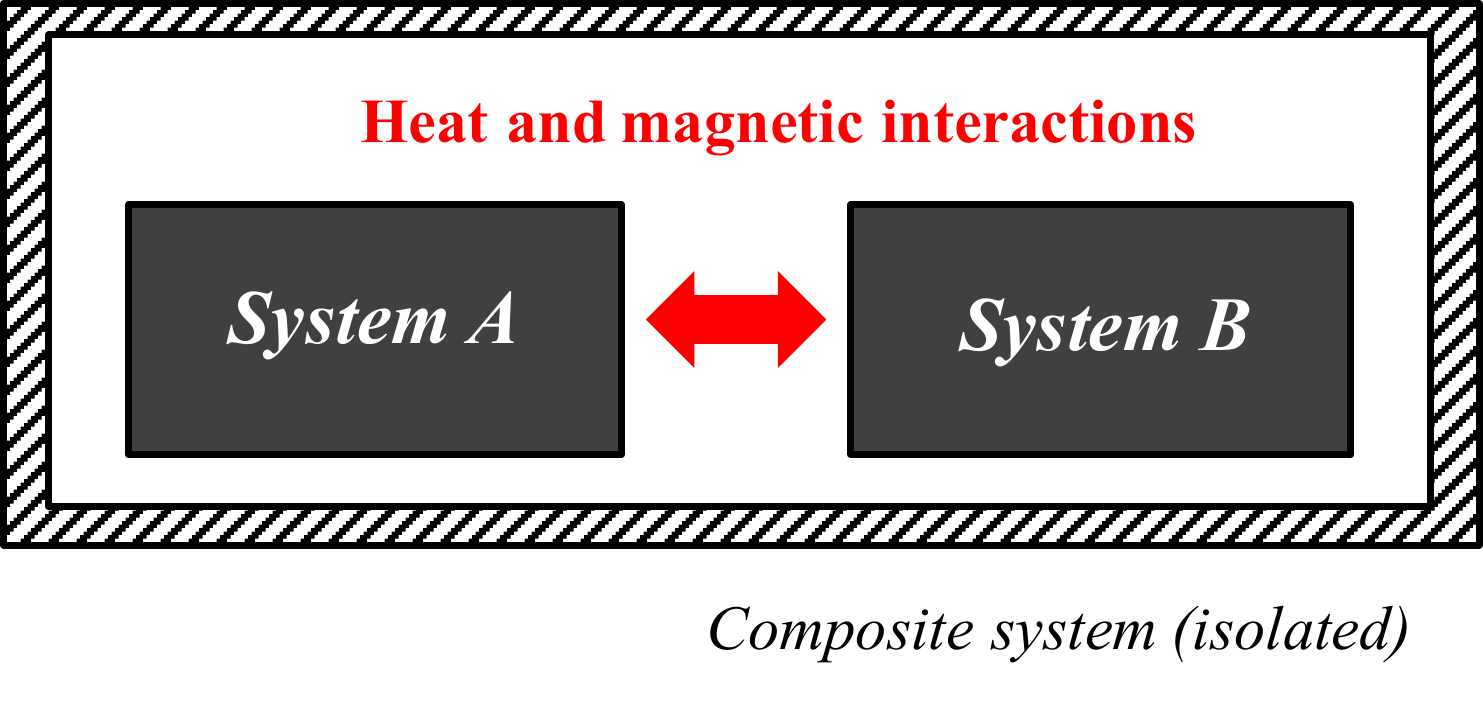}
\caption{\label{fig4:assumed_two_systems} The two interacting systems considered in Sec.\;\ref{chap4_sec:level3_3}. Heat and magnetic interactions between two identical (sub)systems\;$A$ and $B$ are depicted here. Note that the composite system is isolated so that there are no interactions with other systems such as a reservoir.}
\end{center}
\end{figure}

The non-equilibrium temperature and magnetic field strength defined in Sec.\;\ref{chap4_sec:level2_1_3} are fundamental non-equilibrium intensive properties of spin systems that are convenient for analyzing relaxation processes involving not only spin degrees of freedom but lattice and electronic degrees of freedom in a simple way. The use of these non-equilibrium intensive properties is demonstrated in this section by considering heat and magnetic interactions between identical (sub)systems\;$A$ and $B$ (see Fig.\;\ref{fig4:assumed_two_systems}). 

\begin{table}
\begin{center}
\caption{\label{table4:relaxation_patterns_intensive_properties} The initial temperatures and magnetic field strengths of subsystems\;$A $ and $B$ used in the relaxation processes in Sec.\;\ref{chap4_sec:level3_3}. The units of $T_0$ and $H_0$ are, respectively, in K and $\times 10^5$\;Oe. }
\footnotesize
\begin{tabular}{ c  c  c  c  c  c  } 
$\quad$ & $\quad$ & $\quad$ & $\quad$ & $\quad$ & $\quad$ \\ \hline\hline
\quad Process \quad   & \quad  $T^A_0$ \quad  & \quad  $T^B_0$ \quad  & \quad  $H^A_0$ \quad  & \quad  $H^B_0$ \quad \\ \hline
\;1 & 300 & 500 & 0.0 & 0.0   \\
\;2 & 300 & 300 & 0.0 & 1.0  \\
\;3 & 300 & 500 & 0.0 & 1.0   \\ \hline \hline
\end{tabular}
\end{center}
\end{table}

\begin{figure}
\begin{center}
\includegraphics[scale=0.48]{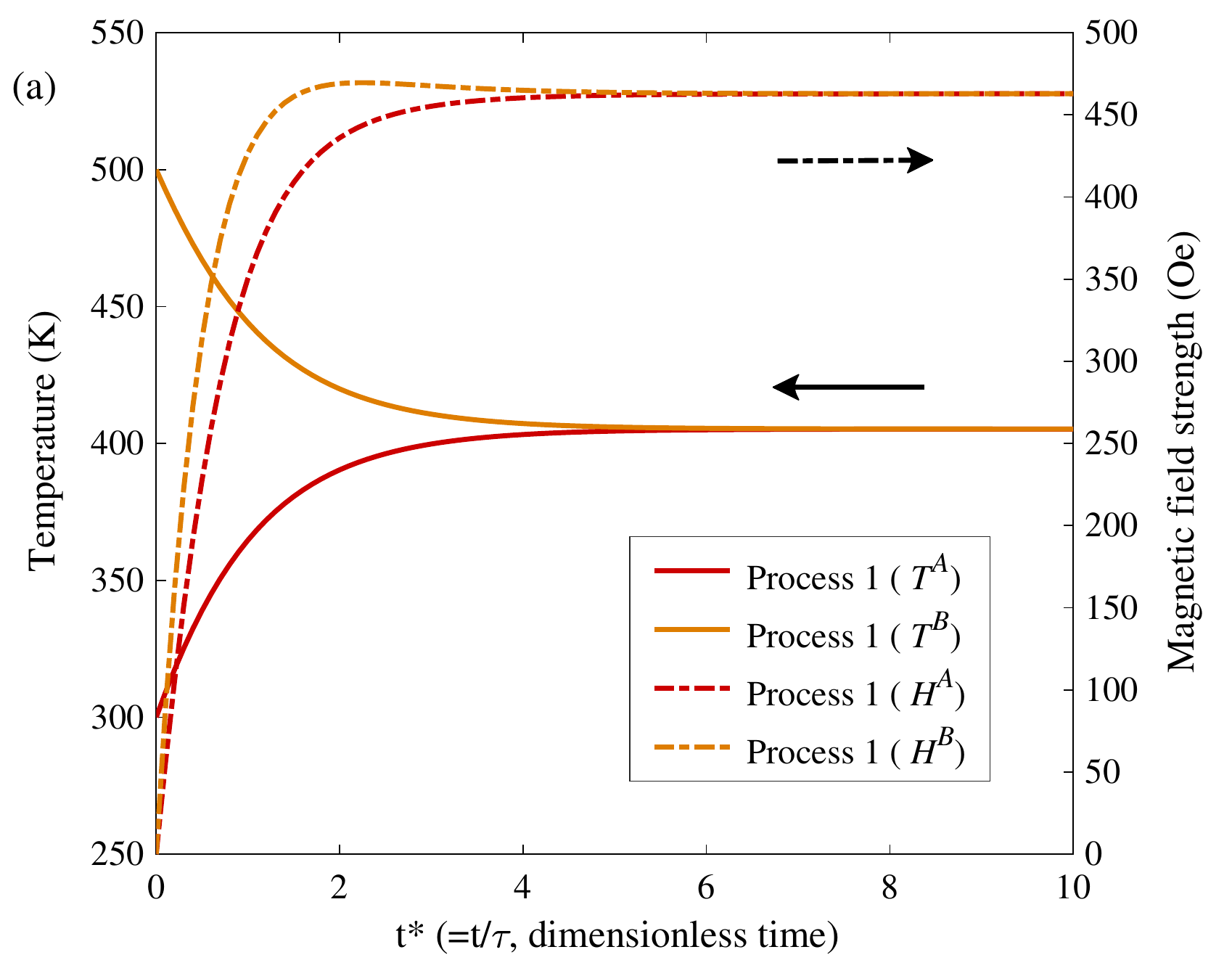}
\includegraphics[scale=0.48]{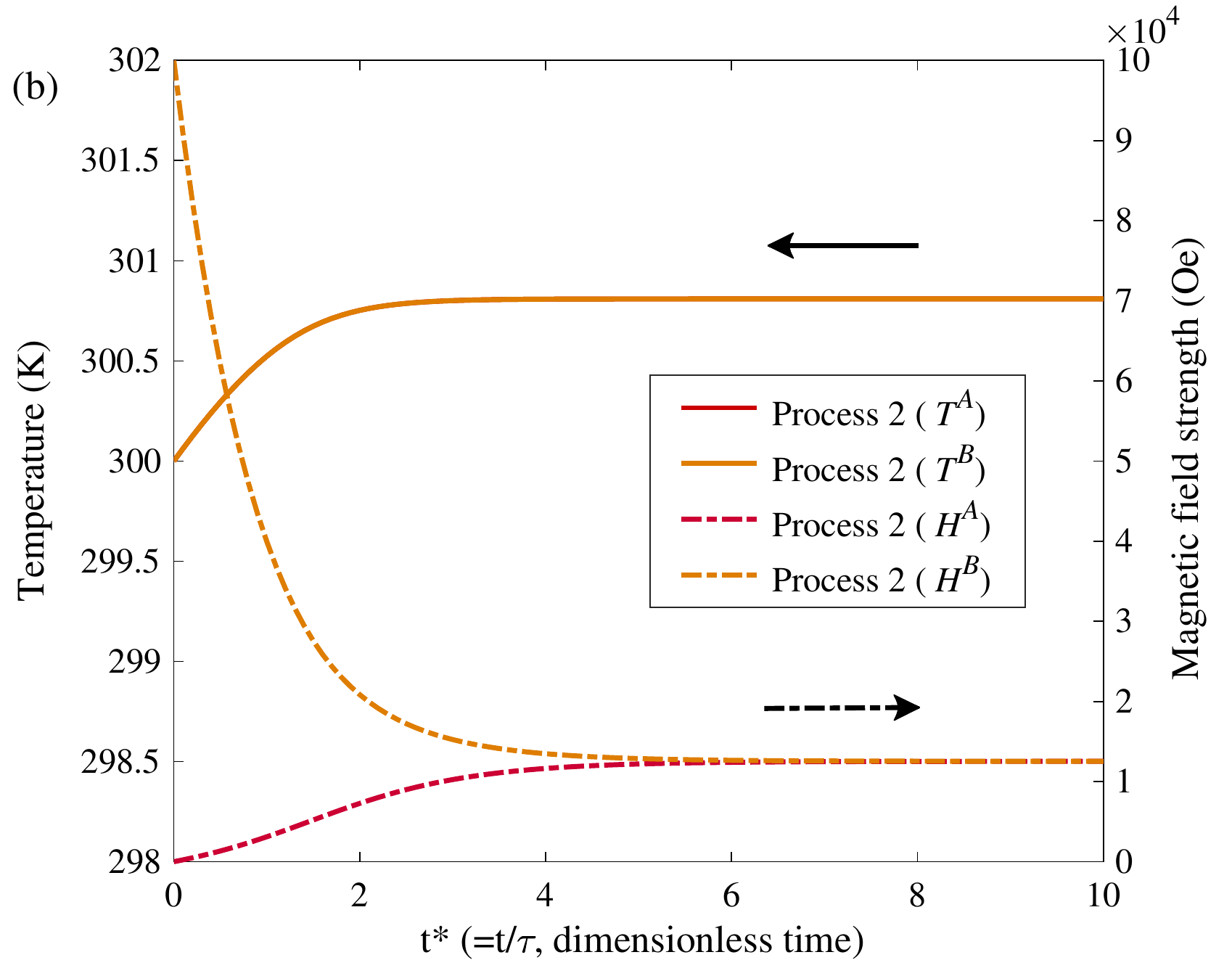}
\includegraphics[scale=0.48]{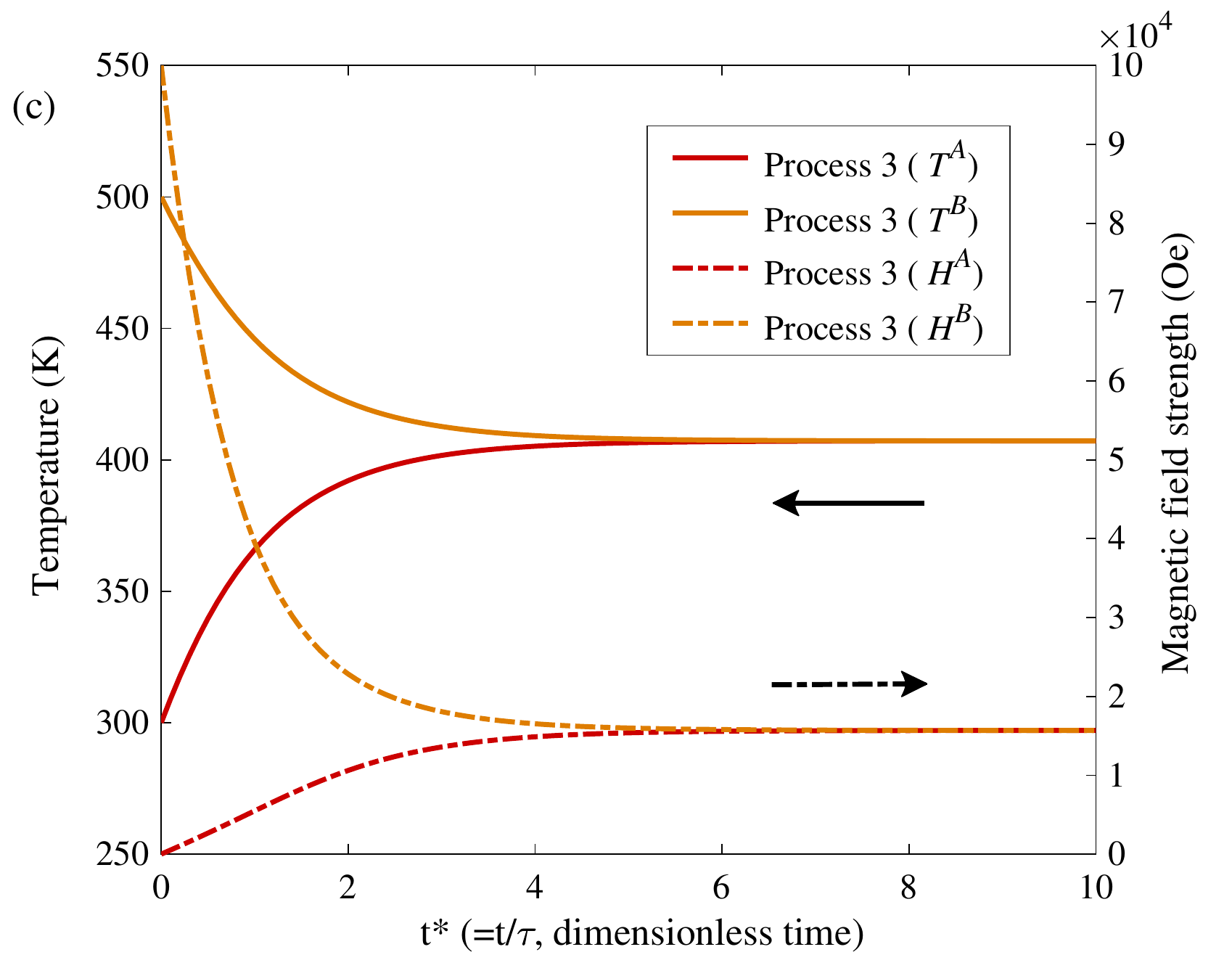}
\caption{\label{fig4:relaxation_process_two_systems_TandH} The calculated time dependence of the intensive properties, temperature and magnetic field strength, of subsystems\;$A$ and $B$ in (a) process\;1, (b) process\;2, and (c) process\;3 (see Table\;\ref{table4:relaxation_patterns_intensive_properties}). The trajectories of temperatures $T^A$ and $T^B$ in process\;2 overlap. The time is normalized by the relaxation time, $\tau$.}
\end{center}
\end{figure}

In order to use the concept of hypoequilibrium states (or non-equilibrium intensive properties), the initial state for the each subsystem is described by a canonical distribution (see Sec.\;\ref{chap4_sec:level2_1_3}) such that 
\begin{equation}
P^0_{k}=\frac{G_{k} e^{- \left( E_{k} - M_{k} H_0 \right) /k_BT_0}}{Z_0} \; ,  \label{eq4:canonical_distribution_initial_state}
\end{equation}
where $Z_0$ is the partition function and the superscripts $A$ or $B$ are omitted. Three different relaxation processes are investigated with the initial temperatures and magnetic field strengths of subsystems\;$A$ and $B$: the temperatures and field strengths of each process are enumerated in Table\;\ref{table4:relaxation_patterns_intensive_properties}. 

\begin{figure}
\begin{center}
\includegraphics[scale=0.5]{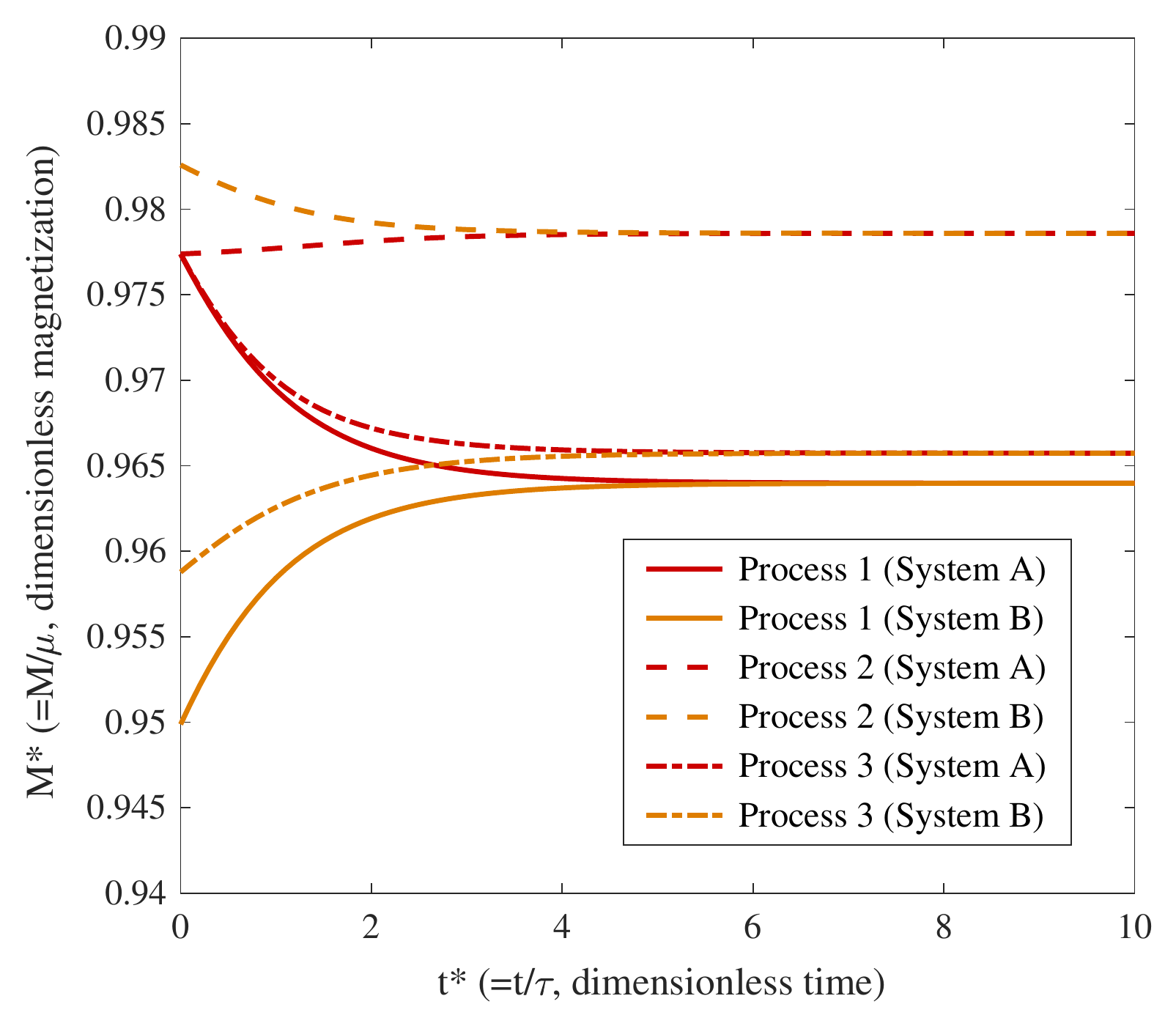}
\caption{\label{fig4:hypo_magnetization_3patterns} The calculated relaxation processes of magnetization in subsystems\;$A$ and $B$. The magnetization is plotted as a fraction of the magnetic moment of iron, $\mu$, and the time is normalized by the relaxation time, $\tau$.}
\end{center}
\end{figure}

The calculated time dependences of intensive properties (temperature and magnetic field strength) and magnetizations of subsystems\;$A$ and $B$ using Eqs.\;(\ref{eq4:second_order_hypoequilibrium_time_evolution}) and (\ref{eq4:equation_of_motion_intensive_properties}) are shown, respectively, in Figs.\;\ref{fig4:relaxation_process_two_systems_TandH} and \ref{fig4:hypo_magnetization_3patterns}. In all of the processes, the final temperatures and magnetic field strengths of subsystems\;$A$ and $B$ converge to the same value (i.e., $T^A=T^B$ and $H^A=H^B$) indicating the subsystems reach mutual equilibrium. In process\;1, the exchange of energy in a heat interaction between the two subsystems produces concomitant changes in the magnetic field strengths as the composite system ($A+B$) evolves to equilibrium. The converse occurs in process\;2: the difference in magnetic field strengths between the two subsystems drives the subsystems to a slightly different temperature.  In process\;3, differences in  both the magnetic field strengths and temperatures of the subsystems produce a subtle interplay between the properties as the composite system relaxes in time. The relaxation behavior of all three processes reflects the magneto-caloric effect in that changes in magnetic interactions between subsystems affect the subsystem temperatures and vice-versa.

\section{\label{chap4_sec:level4}Conclusions}
The SEAQT framework is here used to explore the magnetization of bcc-iron at equilibrium and as it relaxes from non-equilibrium states. The model is based upon the SEAQT equation of motion applied to a pseudo-eigenstructure. Although the energy eigenlevels of the pseudo-eigenstructure could be constructed from any model Hamiltonian, coupled harmonic oscillators are employed here for simplicity. The results confirm that the equilibrium magnetization at low temperatures ($T<500$\;K) with either zero or non-zero external magnetic fields can be estimated reliably with the SEAQT model. They also confirm that the principle of steepest entropy ascent predicts a unique thermodynamic path for the system as it relaxes from an arbitrary, non-equilibrium, initial state (even one far from equilibrium) to stable equilibrium. Furthermore, fundamental non-equilibrium intensive properties (temperature and magnetic field strength) can be defined using the concept of hypoequilibrium states. Relaxation processes in terms of these intensive properties demonstrate the magneto-caloric effect in a straight-forward way.  More complex behavior arising from interactions between magnetic spin and a lattice can be incorporated readily into the pseudo-eigenstructure with approaches analogous to SEAQT models of thermal expansion \cite{yamada2018thermalexpansion} or electron-phonon coupling at an interface \cite{li2018electronphonon}. 


\section*{ACKNOWLEDGEMENT}
We acknowledge the National Science Foundation for financial support through Grant DMR-1506936.

\nocite{apsrev41Control}
\bibliographystyle{apsrev4-2}
\bibliography{ref}

\end{document}